# Breaking the Cascade: Compact Nonlinear Optical Computing with Single-Layer Encoder–Decoder Co-Localization

Yuntian Wang[1,2,3,†], Alexander Chen[1,†], Md Sadman Sakib Rahman[1,2,3], Aydogan Ozcan[1,2,3,*]

[1]Electrical and Computer Engineering Department, University of California, Los Angeles, CA, 90095, USA [2]Bioengineering Department, University of California, Los Angeles, CA, 90095, USA [3]California NanoSystems Institute (CNSI), University of California, Los Angeles, CA, 90095, USA

*ozcan@ucla.edu [†] Equal contribution

## Abstract

Nonlinear computation is fundamental to information processing, yet optical implementations typically rely on either weak material nonlinearities or cascaded architectures that increase system complexity. Here, we demonstrate that nonlinear computing can be achieved with a single linear diffractive surface under coherent illumination. We introduce a compact encoder–decoder co-localization (E+D) architecture in which an input-dependent dynamic encoder and a static optimized decoder are integrated within the same phase-only diffractive plane. Following free-space propagation, coherent interference between the encoder and decoder fields, combined with intensity detection, generates programmable nonlinear input–output mappings without requiring nonlinear optical materials or multiple diffractive layers. We prove that the proposed E+D optical processor is a universal approximator for arbitrary real-valued band-limited nonlinear functions and identify the physical factors governing its approximation fidelity, including the decoder degrees-of-freedom, detector aperture, and axial propagation distance. Crucially, we demonstrate that introducing a trained, frozen phase bias to the encoder region systematically enhances functional expressivity, providing robustness against coarse phase quantization on spatial light modulators. Using this framework, we accurately synthesize diverse nonlinear functions, including commonly used neural network activation functions and complex-valued nonlinear functions. Finally, we experimentally validate the proposed approach using a visible-light optical set-up trained through in situ learning, demonstrating the parallel approximation of 9 nonlinear functions in a single optical forward pass. By collapsing nonlinear optical computation into a single diffractive surface, the E+D architecture substantially reduces hardware and alignment complexity while preserving powerful function-approximation capabilities, providing a compact and scalable framework for analog information processing.

## Introduction

Digital electronic computing has long dominated modern information processing owing to its precision, programmability and scalability. However, the slowdown in conventional device scaling and the rapid growth of machine learning workloads have intensified interest in alternative hardware platforms that can deliver higher computational throughput at lower energy cost[1–4]. Optical computing is attractive in this context because passive wave propagation can manipulate high-dimensional signals with extreme parallelism and low latency[5–7]. Among the various optical computing paradigms, diffractive optical processors have emerged as a versatile paradigm for task-specific analog computing in recent years[8–17]. Architecturally, these systems comprise cascaded, spatially structured passive diffractive surfaces separated from one another by free space. Data-driven, task-specific optimization enables these diffractive processors to manipulate optical fields with high fidelity, successfully approximating various functions using structured, passive linear optical materials[18–20].

Despite these advances, realizing nonlinear mappings in optical hardware, in general, remains challenging due to the weak nonlinearity of optical materials[21–23] and the large intensities required to induce nonlinear interactions[24–26]. Free-space optical computing strategies that rely on nonlinear materials generally require optical intensities that are prohibitive for practical applications[27,28]. Although optical field-enhancement structures can partially mitigate this requirement, they often do so at the cost of introducing additional complexity, optical losses, or bandwidth constraints[29,30]. Furthermore, conventional nonlinear optical processes, such as photochromic or photorefractive effects, typically suffer from slow response times and strong absorption[31–33]. Consequently, overcoming these fundamental limitations without relying on nonlinear optical materials could unlock unprecedented opportunities for scalable, ultrafast, and massively parallel optical computing systems. Partially motivated by these opportunities, recent efforts have explored nonlinear information processing using exclusively linear optical components[26,34–37]. For example, recent work demonstrated massively parallel nonlinear function approximation by encoding the input variable into the phase of a coherent optical wavefront, which is then transformed by an optimized diffractive processor to yield target output fields[18]. Through the interplay between wavefront encoding and spatially varying coherent point spread functions, the framework serves as a universal approximator capable of computing up to one million distinct nonlinear functions in a single forward pass. To accommodate applications demanding varying illumination conditions, this framework was recently expanded to operate under spatially incoherent and partially coherent illumination[38] by employing an intensity-only input encoding. However, existing diffractive architectures for nonlinear function approximation, whether operating under coherent, incoherent or partially coherent illumination, typically rely on multiple diffractive layers or physically separated planes to accommodate input encoding and subsequent wavefront modulation, thereby implementing spatially varying point spread functions programmed on demand. Such an arrangement inherently increases the axial spread of the optical system, requires precise 3D fabrication and/or alignment, and elevates the overall hardware complexity.

Here, we address these limitations by exploring a highly compact optical architecture, termed the encoder–decoder co-localization (E+D) processor, for nonlinear function approximation using a single diffractive surface under coherent illumination. In this framework, one region of the phase-only diffractive surface acts as a dynamic encoder that maps the scalar input variable of the target nonlinear function into an input-dependent phase profile, whereas the remaining region serves as an optimized static decoder. Because the encoder and decoder are co-localized on the same diffractive plane, the architecture eliminates the need for multiple cascaded layers or physically separated modulation stages, thereby reducing both the optical processor volume and the alignment complexity of the experimental set-up. After free-space propagation, the input-dependent encoder field and the structured decoder field interfere at the output plane; intensity detection converts this coherent superposition into a real-valued signal containing the mixing term required for nonlinear function synthesis. Building on this mechanism, we theoretically and numerically show that the single-layer co-localized E+D processor can approximate arbitrary real-valued band-limited nonlinear functions. We further demonstrate the use of the E+D processor to approximate commonly used nonlinear activation functions, e.g., Rectified Linear Unit, Sigmoid, Exponential Linear Unit, and Gaussian Error Linear Unit, as well as complex-valued nonlinear functions. We also identify the key architectural factors that govern its approximation fidelity and show that a trained, frozen encoder bias improves both function approximation accuracy and robustness to finite-phase quantization at the input plane. Beyond theoretical and numerical analyses, we further validate this architecture experimentally using a visible-light optical set-up, where the E+D processor is trained on hardware through model-free in situ learning to simultaneously implement 9 distinct nonlinear functions in a single pass through the optical processor. Together, these results establish a compact, easy-to-align and physically interpretable framework for scalable nonlinear analog optical computing.

## Results

We illustrate the operating principles of the E+D processor in **Fig. 1**. As shown in **Fig. 1a**, the system consists of a single phase-only diffractive surface that is partitioned into a dynamic encoder region that represents the input function arguments and a static decoder region, co-localized on the same input plane. Under coherent plane wave illumination, the encoder maps the scalar input $a$ of the target nonlinear function $f(a)$ into an input-dependent phase profile $\varphi(p; a) = -2\pi pa$, where $p$ denotes the spatial coordinate/index on the input plane. The static decoder phase pattern, parameterized by the trainable phase distribution $\theta$, remains fixed once optimized to synthesize the target nonlinear function. Following free-space propagation, the resulting optical field is converted into a scalar response that follows $f(a)$ through single-pixel intensity detection at the output plane; see **Figure 1b**. After training, the detected output, for any input $a$, closely matches the target nonlinear function over the input range of interest (defined as $a \in [0, 1]$ throughout this work unless otherwise specified). In the next subsection, we theoretically establish the feasibility of nonlinear function synthesis via co-localization of the argument-dependent encoder phase and a static decoder, all within a single linear diffractive surface, followed by intensity detection.

### Theoretical analysis

We consider a diffractive optical processor comprising a single diffractive surface, operating under coherent illumination. For mathematical clarity, a one-dimensional formulation is presented, and the extension to higher-dimensional nonlinear functions is straightforward. The diffractive surface is assumed to be located at $z = 0$, and an intensity detector is placed at the output plane $z = z_0$. The diffractive surface is partitioned into (i) a dynamic encoder region whose phase profile encodes the input variable $a$ of the target function, and (ii) a static decoder region whose optimized phase profile remains fixed, i.e., the decoder phase profile does not depend on $a$. Therefore, we term the proposed architecture a co-localized E+D processor.

Within the encoder region ($|x| \leq E$), the input variable $a$ is encoded as,

$$\varphi_{\text{enc}}(x; a) = -2\pi a x. \tag{1}$$

Here $2E$ denotes the width of the encoder region. Let $h(\cdot)$ denote the coherent point spread function between the diffractive E+D surface and the output plane; by definition $h(\cdot)$ is an even function. The encoder-generated field at the output plane coordinate $u$ is then given by,

$$E_{\text{enc}}(u; a) = \int_{-E}^{E} h(u - x) e^{j\varphi_{\text{enc}}(x;a)}\, \mathrm{d}x = e^{-j2\pi a u} \int_{u-E}^{u+E} h(s)\ e^{j2\pi a s}\, \mathrm{d}s = e^{-j2\pi a u} H_E(u; a), \tag{2}$$

where $H_E(u; a) = \int_{u-E}^{u+E} h(s)\ e^{j2\pi a s}\, \mathrm{d}s$ and $h(s) = h(-s)$. When the output detector spans a small window $\mathcal{W} = \{u: |u| \leq w/2\}$ with a full width of $w$ and the encoder aperture is sufficiently large compared to both the effective support of $h$ and the detector window (i.e., for large $E$ and/or small $z_0$, yielding $E \gg w/2$), $H_E(u; a)$ can be simplified as the free-space transfer function $H(a)$,

$$H_E(u; a) \approx H(a) = \int_{-\infty}^{\infty} h(s)\, e^{-j2\pi a s}\, \mathrm{d}s, \tag{3}$$

which yields:

$$E_{\text{enc}}(u; a) \approx e^{-j2\pi a u}\, H(a). \tag{4}$$

Let us denote the static decoder region induced complex field at the output plane as $E_{\text{dec}}(u)$, then, the single pixel detector intensity over the region $\mathcal{W}$ can be written as:

$$I(a) = \int_{-w/2}^{w/2} |E_{\text{enc}}(u; a) + E_{\text{dec}}(u)|^2\, \mathrm{d}u. \tag{5}$$

Using Eq. (4), $I(a)$ can be expressed as:

$$I(a) = \int_{-w/2}^{w/2} |H(a)|^2 \,\mathrm{d}u + \int_{-w/2}^{w/2} |E_{\mathrm{dec}}(u)|^2 \,\mathrm{d}u + \int_{-w/2}^{w/2} 2\Re\{e^{-j2\pi a u}\, H(a) E_{\mathrm{dec}}^*(u)\} \,\mathrm{d}u$$
$$= w|H(a)|^2 + \beta + 2\Re\{H(a)C(a)\}, \quad (6)$$

where

$$\beta = \int_{-w/2}^{w/2} |E_{\mathrm{dec}}(u)|^2 \,\mathrm{d}u, \quad (7)$$

and

$$C(a) = \int_{-w/2}^{w/2} E_{\mathrm{dec}}^*(u)\, e^{-j2\pi a u} \,\mathrm{d}u. \quad (8)$$

Eq. (6) reveals the physical origin of the nonlinear function approximation capability of an E+D processor. Although propagation through a diffractive surface is linear in the complex field, the output detector converts the coherent superposition into an intensity signal containing the multiplicative mixing term $2\Re\{H(a)C(a)\}$ between the input-dependent encoder field and the optimized decoder field. This interference-induced mixing is what enables *nonlinear* function synthesis. To synthesize any arbitrary real-valued nonlinear band-limited function $f(a)$, we consider the condition that the effective interference response $C(a)$ is programmed (i.e., optimized through the degrees of freedom of the decoder pixels) to satisfy:

$$C(a) = H^*(a)\left[-w/2 + \frac{1}{2} f(a) + j g(a)\right], \quad (9)$$

where $g(a)$ is an arbitrary real-valued auxiliary function. Substituting Eq. (9) into Eq. (6) gives

$$I(a) = \beta + |H(a)|^2 f(a). \quad (10)$$

Because $\beta$ is a constant, independent of the input encoding $a$, it can be removed by background subtraction. Under free-space propagation, $|H(a)|^2 = 1$ within the propagating spatial-frequency band (see Methods for details). Therefore, the ideal detector response after the background removal is proportional to $f(a)$.

Due to the arbitrary selection of $g(a)$), Eq. (9) specifies a set of non-unique solutions to $C(a)$, which can be synthesized through the optimization of the degrees of freedom of the decoder. Let $\varphi_{\mathrm{dec}}(x)$ denote the static and optimized phase profile of the decoder region for optically implementing the nonlinear function $f(a)$. The physically realizable decoder field at the output plane can be written as,

$$E_{\mathrm{dec}}(u) = \int_{\mathcal{D}} h(u - x) \mathrm{e}^{j\varphi_{\mathrm{dec}}(x)} \, \mathrm{d}x, \tag{11}$$

where $\mathcal{D}$ covers an arbitrary design map, denoting the decoder-assigned pixels at the E+D processor surface. Mapping $C(a)$ to the ideal set of functions defined in Eq. (9) through a physical decoder phase profile, $\varphi_{\mathrm{dec}}(x)$, constitutes a non-unique inverse problem since many solutions can provide similar approximations. In practice, rather than attempting to analytically derive a suitable decoder phase profile $\varphi_{\mathrm{dec}}(x)$ that satisfies Eq. (9), here we optimize it directly via a differentiable, physics-aware digital twin using gradient-based numerical methods, as detailed in the Methods section. Within this framework, the auxiliary function $g(a)$ in Eq. (9) represents additional degrees of freedom allowed by intensity-only detection, thereby enlarging the solution space for physically realizable phase-only decoders with $\varphi_{\mathrm{dec}}(x)$.

We initially demonstrate the feasibility of this E+D processor analysis through numerical simulations (**Figs. 2-3**), where we used a spatially coherent illumination at a wavelength of $\lambda =$ 660 nm and a simulation pixel pitch of 300 nm. The encoder region was selected as a square block of $N_p$ pixels ($\sqrt{N_p}$ pixels along each side), placed at the center of the input E+D plane, where $N_p$ denotes the number of harmonics of the target real-valued nonlinear function, $f(a)$, and the surrounding pixels at the E+D plane form the decoder region ($N_d$ pixels in total, assigned to the fixed/optimized decoder, defined by $\mathcal{D}$). The scalar input $a$ is drawn from the range $[0, 1]$, and the propagation distance is fixed at $dz = 10\lambda$ unless otherwise stated. In our initial numerical analysis, the target functions are represented using $N_p = 16$ Fourier harmonics and each $(w, N_d)$ data point in **Figs. 2-3** reports the mean and standard deviation across 20 independently trained/optimized E+D processor models.

Two design parameters are central to the approximation capabilities of the presented E+D processor: the detector width $w$, which determines how much of the output interference pattern is collected, and the decoder size $N_d$, which sets the number of trainable phase elements (degrees of freedom) available for shaping the nonlinear response of the optical processor. Our analysis in **Fig. 2-3** reveals that, at a moderately small detector width ($w = 2\lambda$), increasing the decoder degrees of freedom ($N_d$) reduces the MSE (mean squared error) by several orders of magnitude, from $\sim 10^{-2}$ to $\sim 10^{-8}$. However, at a larger detector width ($w = 4\lambda$), this $N_d$-dependent trend largely saturates: all evaluated decoder sizes achieve exceptionally low MSE, falling below the precision limit of the simulations. In this regime, the E+D processor approaches the idealized response described in Eqs. (4)–(9), wherein the target nonlinear function synthesis condition reduces to realizing the band-limited interference cross-term $C(a)$. Beyond the minimum $N_d$ required to span the target decoder family of functions, set by Eq. (9), additional decoder elements no longer enlarge the relevant solution space.

Taken together, these simulation results confirm our theoretical analysis that accurate nonlinear function synthesis necessitates both an adequately large detector aperture and sufficiently large decoder capacity ($N_d$). The same conclusions can also be drawn for the approximation of nonlinear functions with $N_p = 9$ harmonics, reported in **Supplementary Fig. S1**.

**Figures 2-3** also cover the case when $w$ is significantly small, i.e., the output detector no longer integrates a spatially extended interference pattern; instead, it probes a highly localized region of the propagated field. In this limit, the collected output intensity in Eq. (6) reduces to,

$$I(a) \approx w + \beta + 2\Re\{H(a)E_{\mathrm{dec}}^{*}(0)\}, \quad (12)$$

This demonstrates that, as $w \to \lambda/2$, the decoder modulates the output signal intensity exclusively through its local complex value at the detector position, rather than through a spatially distributed interference pattern. Consequently, the capacity to synthesize a nonlinear response is severely restricted, yielding persistently high approximation errors across all tested $N_d$ values at small detector sizes (**Fig. 2a**). The representative examples illustrated in **Fig. 2b** further elucidate this failure mechanism: when the detector aperture is too small to capture the relevant spatial structure of the mixed field, the diffractive output no longer reproduces the target function profile.

Beyond the detector width $w$ and decoder dimensionality $N_d$, the approximation fidelity of the coherent E+D processor is also influenced by the free-space propagation distance $dz$ between the diffractive surface and the output plane, as shown in **Fig. 3**. Physically, $dz$ and $w$ play complementary roles in governing the extent to which the encoder and decoder fields interfere within the detector window. A longer propagation distance allows diffraction to laterally spread both fields across the output plane, thereby increasing their spatial overlap across a given detector aperture and propelling the system toward the idealized large-aperture regime described by Eqs. (3)–(4). Accordingly, at a moderate detector width ($w = 2\lambda$), increasing $dz$ markedly lowers the nonlinear function approximation errors for both $N_d = 3{,}200$ (**Fig. 3a**) and $N_d = 6{,}400$ (**Fig. 3b**), an observation consistent with more complete coherent mixing within the detector region. However, at a larger detector width ($w = 4\lambda$), this $dz$-dependent trend largely saturates. Because the detector is already sufficiently wide to collect the majority of the mixed interference pattern, the system effectively reaches the idealized response regime, meaning that further increases in $dz$ no longer expand the usable response space. In this saturated regime, the residual performance differences among various $dz$ values are statistically negligible and, consistent with the findings in **Fig. 2**, they are dominated by the variance inherent to stochastic gradient-based optimization rather than by any fundamental shift in representational capacity of the E+D processor.

### Improved nonlinear function approximation through a trained encoder bias

To further enhance the approximation capability of the coherent E+D processor without altering its single-layer geometry, we next augment the argument-dependent encoder with an additive trained phase bias that is frozen, as illustrated in **Fig. 4a**. Specifically, the encoder phase profile is modified to $\varphi_{\text{enc}}(x;a) = -2\pi ax + b(x)$, where $b(x)$ is a spatially varying phase bias that is optimizable during training and frozen at inference. This modification preserves the co-localized encoder–decoder architecture, and introduces additional trainable degrees of freedom within the encoder region, enhancing the expressivity of the encoder fields that interfere with the decoder field at the output plane.

We evaluated the impact of $b(x)$ statistically over 40 independently sampled target nonlinear functions, each implemented with a separately trained E+D processor model. The input plane comprises $42 \times 42$ pixels, with a central $4 \times 4$ encoder block ($N_p = 16$) and the remaining 1,748 pixels forming the decoder region. The single-pixel output detector spans a $2\lambda \times 2\lambda$ region at the output plane. The analyses reported in **Fig. 4** yield two primary observations. First, the trained encoder bias systematically shifts the MSE distribution toward lower values, as illustrated in **Fig. 4b**, demonstrating a consistent and major improvement in the nonlinear function approximation fidelity of the E+D processor. Second, the best-, median-, and worst-case examples depicted in **Fig. 4c** exhibit tighter agreement with the target nonlinear functions under the bias-augmented E+D processor designs. The better function-approximation results indicate that the performance gain is robust to varying degrees of approximation difficulty, rather than confined to trivially approximated targets. Notably, this enhancement is most pronounced in the worst-case scenario, where the baseline encoder fails to accurately reconstruct the target function—precisely the regime where the additional degrees of freedom provided by the bias-augmented encoder are most critical.

Having established the statistical advantages of the trained encoder bias in the E+D processor architecture, the superiority in performance was further validated on four representative nonlinear activation functions (i.e., Rectified Linear Unit[39] (ReLU), Sigmoid[40], Exponential Linear Unit[41] (ELU), and Gaussian Error Linear Unit[42] (GELU)) widely used in modern neural networks, as shown in **Fig. 5a**. For this study, the input plane comprises $70 \times 70$ pixels, with a central $8 \times 8$ encoder block ($N_p = 64$) surrounded by 4,836 decoder pixels. The detector covers a $10\lambda \times 10\lambda$ region, the propagation distance is $dz = 100\lambda$ to ensure adequate interference between the encoder and decoder induced fields, and the scalar input $a$ is drawn from $[-0.5, 0.5]$ so that the selected nonlinear activation functions are evaluated symmetrically across their transition regions. Across all four distinct activation functions, the bias-augmented E+D designs yielded closer agreement with the ground truth than the bias-free baseline processor designs, with the most visible improvements occurring in regions of rapid transitions or pronounced curvature, as highlighted by the enlarged views in **Fig. 5**. These results indicate that the learned encoder bias improves not only average performance over randomly selected target function ensembles, but also the approximation of structured nonlinear functions of practical interest.

To visualize how the learned bias term reshapes the processor's learned phase profile, **Fig. 5b** shows the learned phase profiles of the E+D processor at a representative input $a = 0.5$ for each of the four target nonlinear functions. In the bias-free baselines (left column), the encoder region is identical across all four targets, since $\varphi_{\text{enc}}(x;a) = -2\pi ax$ carries no function-specific information, and all adaptation must occur in the static decoder region. In the bias-augmented designs (right column), the same $a$ yields different encoder profiles across the four targets, reflecting the target-specific static bias $b(x)$ learned during training. The decoder region remains the dominant repository of function-specific phase structure in both cases, but the bias-augmented design additionally exploits the encoder region as a tunable spatial channel, enlarging the accessible solution space of the E+D processor.

We further investigated whether this significant advantage provided by the trained encoder bias persists when the input-plane phase is quantized. This analysis is of direct experimental relevance since spatial light modulators (SLMs), the natural hardware platform for implementing the E+D processor, impose a finite phase bit depth on the input plane, and reliable operation under this constraint is a prerequisite for any physical realization of the E+D processor. In this phase bit-depth analysis, the input plane comprises $40 \times 40$ pixels, with a central $4 \times 4$ encoder block ($N_p = 16$) surrounded by 1,584 decoder pixels and a single pixel output detector spanning $2\lambda \times 2\lambda$. For each quantization bit depth, statistics are compiled over 40 independently sampled target nonlinear functions, each optimized with a separately trained E+D processor model. Across all tested quantization levels, the bias-augmented encoders consistently yield lower error distributions than the bias-free baselines, as shown in **Fig. 6a**, indicating that the phase-bias-induced performance gain is retained under the hardware-relevant phase bit-depth constraint of coarse-phase quantization at the input plane. Representative approximations in **Fig. 6b** further show that, at a given phase quantization level, the trained encoder bias enables a closer match to the target function and better preserves the overall nonlinear function profile at the output.

## Complex-valued nonlinear function approximation using E+D processors

The presented single surface-based E+D processor architecture cannot be used for complex-valued nonlinear function approximation using the output complex optical field. This is a limitation of the E+D processor architecture compared to multi-layered diffractive processors that were demonstrated to approximate cascadable (field-based) complex-valued nonlinear functions at their outputs[18]. If intensity detection is not performed at the output plane of the E+D processor, the signal mixing between the encoder and decoder fields in Eqs. (5-6) will be replaced by a simple linear superposition of the encoder and decoder complex fields, which limits the function representation power of an E+D processor in the complex field domain. These limitations, however, do not apply to deeper, multi-layer diffractive processors that can synthesize multi-dimensional complex-valued nonlinear functions at their output fields[18], which constitute a super-set to E+D processors.

Although an E+D processor cannot directly synthesize at its output field an all-optically cascadable complex-valued nonlinear function, it can still be programmed to synthesize any arbitrary complex-valued nonlinear function by using 2 separated intensity-only output detectors each of which is assigned to a distinct real-valued nonlinear function (corresponding to the real and imaginary parts of the target complex-valued nonlinear function), which will be demonstrated in this section (see **Fig. 7**). Before, we showcase this capability, we would like to further shed light on the complex field processing capabilities of the E+D processor architecture, and consider here a Gedanken control experiment in which we hypothetically measure the real part of the propagated field at the detection/analysis plane,

$$\hat{f}_{real}(a) = \int_{-w/2}^{w/2} \Re\{E_{\text{enc}}(u;a) + E_{\text{dec}}(u)\} \, du. \tag{13}$$

Substituting the large-aperture approximation of Eq. (4) into the real valued readout yields:

$$\hat{f}_{real}(a) \approx w\Re\{H(a)\text{sinc}(\pi a w)\} + \Re\left\{\int_{-w/2}^{w/2} E_{\text{dec}}(u) \, du\right\}, \tag{14}$$

where $\text{sinc}(x) = \sin x / x$. This reveals that the second term in Eq. (14) is a constant, independent of $a$, and the first term represents an input-dependent encoder field modulated by a sinc envelope fixed by the detector geometry. Crucially, Eq. (14) contains no multiplicative mixing term between the encoder field and the decoder field like the quadratic cross-term $2\Re\{H(a)C(a)\}$ present in Eq. (6), which is precisely the term that carries the synthesizable nonlinear response in the earlier results that we presented. Therefore, the trainable decoder in this case is restricted to applying a uniform baseline shift, leaving it incapable of modulating the system's functional dependence on the input variable $a$. These predictions are also confirmed numerically in **Supplementary Fig.S2a-b**. Under the real-valued linear-readout configuration, the approximation error remains high across all the tested detector widths $w$ and decoder feature numbers $N_d$, and the predicted outputs fail to reproduce the target profiles. A similar Gedanken configuration utilizing the imaginary part (instead of the real part introduced in Eq. (13)) reveal the same conclusions, reported in **Supplementary Fig.S3**.

These analyses confirm that the presented E+D processor architecture cannot be used for complex-valued nonlinear function approximation in the output field. Despite its simpler architecture involving a single SLM plane where the encoder and decoder structures spatially overlap, it cannot be used for synthesizing optically cascadable nonlinear functions using complex fields[18]. A similar limitation occurs in the case of spatially incoherent illumination even if an intensity detector is used at the output of an E+D processor, which is discussed in the next subsection. These limitations can be overcome by deeper diffractive processors that can approximate multi-dimensional complex-valued nonlinear functions at their output fields[18].

While the E+D processors cannot directly synthesize optically cascadable complex-valued nonlinear functions in their output fields, they can be used to represent any arbitrary complex-valued nonlinear function by using 2 separated intensity-only output detectors each of which is assigned to a distinct real-valued nonlinear function, $f_k(a)$, $k = 0, 1$ (see **Fig. 7**). By pre-assigning each real-valued nonlinear function at each output detector of the E+D processor to a complex valued basis $e_k = \exp\left(jk\frac{\pi}{2}\right)$, any arbitrary complex-valued bandlimited nonlinear function can be synthesized by $f(a) = \sum_{k=0}^{1} e_k f_k(a)$ . **Figure 7a** validates this two-detector scheme numerically: the left panels show the E+D processor output plane layout with two distinct output detectors (spanning $4\lambda \times 4\lambda$ with a gap of $10\lambda$ in between) and the corresponding complex-valued bases, the center panels display the real parts of the individually approximated nonlinear functions $f_{0,}(a)$ and $f_1(a)$, and the right panel overlays the reconstructed complex-valued function trajectory in the complex coordinate against the ground-truth target nonlinear function, confirming close agreement between the two. The optimized decoder and encoder bias overlay is also shown in **Fig. 7a**.

If $f_k(a)$ is restricted to real-valued non-negative nonlinear functions (eliminating the background removal step in Eq. (10)), then 3 output detectors, each assigned to a distinct real-valued non-negative nonlinear function, $f_k(a)$, $k$ =0, 1, 2 would be sufficient to approximate any arbitrary complex-valued band-limited nonlinear function at the output, which can be synthesized by $f(a) = \sum_{k=0}^{2} e_k f_k(a)$ where $e_k = \exp\left(jk\frac{2\pi}{3}\right)$. **Figure 7b** demonstrates this three-detector configuration: with the detector size spanning $4\lambda \times 4\lambda$ and $7.5\lambda$ gap between detectors, the three base functions are each approximated by a dedicated optical detector and the resulting complex-valued reconstruction again closely tracks the target nonlinear function in the complex plane. The two-detector configuration achieves MSE values on the order of $\sim 10^{-8}$, while the more constrained three-detector configuration, which enforces non-negative real-valued component functions, attains MSE values on the order of $\sim 10^{-3}$. Both results confirm that the complex-valued function synthesis maintains high approximation fidelity consistent with the real-valued case.

**Spatially incoherent E+D processors**

As detailed in this section, under spatially incoherent illumination a single trainable diffractive layer within the E+D processor architecture cannot approximate an arbitrary non-negative nonlinear function. To analyze the capabilities of an E+D processor under spatially incoherent illumination, we follow a similar strategy. We hold the same assumptions made earlier for the spatially coherent case, regarding the input and output planes in terms of distances and encoder/decoder regions. In this spatially incoherent case, without loss of generality, we choose our encoder function as an intensity modulation $[\cos(2\pi a x) + 1]$, where $x$ represents the coordinates of the input encoder region. Assume that the intensity point spread function between the input and output planes is represented by a non-negative real-valued function $h_i(u)$, where $u$ represents the coordinates of the output plane. The encoder-created intensity pattern, $E_i(u)$, at the output plane can be written as:

$$E_i(u) = \int_{-E}^{E} [\cos(2\pi a x) + 1] \; h_i(u - x) dx \quad (15)$$

where $[-E: E]$ defines the intensity encoder region at the input plane. Assuming the encoder region is significantly large with respect to the effective width of $h_i$ and the output detector region, which can be satisfied by appropriately decreasing $z_0$ and increasing $E$, one can write the encoder-created output intensity pattern as:

$$E_i(u) = H_i(0) + Real\{H_i(a) e^{-j2\pi u a}\} \quad (16)$$

where $H_i$ is the Fourier Transform of $h_i$. Similar to earlier sections, we assume that the rest of the input plane is dedicated to an optimizable decoder, where the transmission values of these decoder regions at the input are fixed after optimization and are independent of the function argument, $a$. Therefore, this optimizable decoder transmission structure at the input plane will create a decoder intensity pattern at the output, given by $D_i(u)$. For an opto-electronic detector centered at $u = 0$, with a full width of $w$, the resulting output optical signal, $I$, for each value of the function argument can be written as:

$$I = \int_{-w/2}^{w/2} [E_i(u) + D_i(u)]du$$
$$= wH_i(0) + \alpha + Real\left\{\int_{-w/2}^{w/2} H_i(a)e^{-j2\pi ua}\, du\right\}$$
$$= wH_i(0) + \alpha + H_i(a)K_i(a) \quad (17)$$

where $\alpha = \int_{-w/2}^{w/2} D_i(u)\, du$ is a constant independent of $a$, and $K_i(a) = \int_{-w/2}^{w/2} \cos(2\pi ua)du$. If we assume $w$ is large enough, $K_i(a) \approx \delta(a)$, which yields, $I \approx [w + \delta(a)]H_i(0) + \alpha$.

Unlike Eqs. (5-6) of the spatially coherent illumination case, in the spatially incoherent illumination, there is no signal mixing between the encoder and decoder patterns even if an intensity detector is used at the output since their intensities linearly add up at the detector area. Therefore, due to the highly restricted functional form of the output signal $I$ under spatially incoherent illumination (Eq. (17)), a single trainable diffractive layer with the E+D processor architecture cannot approximate an arbitrary non-negative nonlinear function of interest. For spatially incoherent illumination, a deeper diffractive architecture with multiple layers can be used to approximate an arbitrary bandlimited nonlinear function[38], including complex-valued nonlinear functions. These conclusions further support the mathematical advantages of structural depth in diffractive optical processors.

**Experimental demonstration of spatially-multiplexed nonlinear-function approximation using a coherent E+D processor**

We experimentally demonstrated a coherent E+D processor for parallel nonlinear function approximation using a visible-light set-up. In this proof-of-concept experiment, a coherent laser beam at a wavelength of 635 nm is modulated by an 8-bit phase-only SLM with an 8 $\mu m$ pixel pitch and recorded by a CMOS sensor after propagation through a 4f relay system with the schematic shown in **Fig. 8a**; also see **Supplementary Fig. S4**. The 4f system is used to relay the optical field and adjust the effective propagation distance. To simultaneously implement nine randomly selected nonlinear functions, the corresponding E+D units are spatially multiplexed by arranging them in a 3×3 mosaic on the SLM, with nine associated output detector regions (one for each nonlinear function) defined on the camera plane (see Methods section for details). For each input argument $a$, the E+D processor phase profile was constructed by adding the argument dependent encoder region to the optimized decoder region, including the trained phase bias during inference (as shown in **Fig. 8b**).

To account for optical misalignments, phase-response nonidealities, and measurement noise, the models were first trained *in silico* using a digital twin and subsequently fine-tuned directly on the optical hardware utilizing a model-free *in situ* optimization procedure. Specifically, we formulated this hardware fine-tuning process as a continuous-space reinforcement learning problem and employed a Proximal Policy Optimization (PPO) algorithm[43]. Within this framework, the phase value of each individual pixel—spanning both the decoder region and the optimizable encoder bias—was modeled as an independent Gaussian stochastic policy. Each pixel was parameterized by an optimizable mean ($\mu$) to steer the policy updates and a fixed standard deviation ($\sigma$) to control the exploration variance. At each training epoch, 64 candidate phase profiles were sampled from these pixel-wise distributions and experimentally evaluated using randomly sampled 16-point batches of the input argument $a$. The candidate profile achieving the lowest training MSE, which served as the primary reward function (minus MSE) for the PPO algorithm, was subsequently evaluated on a fixed, uniformly sampled 51-point grid of $a$ values, yielding the evaluation MSE reported throughout the training process (shown in **Fig. 8c**). As the PPO-driven *in situ* optimization proceeded, the experimentally measured outputs progressively converged toward the target nonlinear functions, with the best epoch reaching a mean evaluation MSE of $\sim 10^{-2}$ across the nine simultaneously approximated nonlinear functions. The nonlinear function approximation results of epoch 0, 300 and 660 during the in-situ learning process are reported in **Fig. 8d**.

These results experimentally validate that a coherent E+D processor can physically implement multiple nonlinear functions, all in parallel, through the interplay of phase modulation, free-space propagation, and output intensity detection. Also see **Supplementary Fig. S5** for additional experimental results further supporting our conclusions.

## Discussion

In this work, we introduced the E+D optical processor, a single-surface diffractive architecture that achieves nonlinear function approximation under coherent illumination using only linear optical materials. By co-localizing a dynamic input encoder and a static optimized decoder on the same phase-only diffractive plane (e.g., a single SLM) and reading out the propagated field with an intensity detector, the architecture realizes a universal approximator for real-valued band-limited nonlinear functions without cascaded diffractive elements or nonlinear optical materials. Compared with previously reported coherent[18] and incoherent[38] diffractive networks for nonlinear function approximation, which rely on multiple axially separated diffractive layers, the E+D optical processor collapses these functions onto a single SLM plane. This reduction in architectural depth eliminates the necessity for inter-layer alignment, shrinks the axial footprint between the diffractive surface and the detector, and, importantly, does so without sacrificing approximation capacity — mean squared errors as low as $\sim 10^{-12}$ can be reached numerically once the detector aperture, decoder dimensionality, and propagation distance are optimized, as demonstrated in the Results section. While possessing these important advantages, its shallow architecture also limits the capabilities an E+D processor as reported in the previous sub-sections: (1) an E+D processor cannot directly synthesize optically cascadable complex-valued nonlinear functions in its output fields, and (2) under spatially incoherent illumination, an E+D processor cannot approximate an arbitrary non-negative nonlinear function of interest. Both of these are attainable using deeper, multi-layer diffractive optical processors[18,38], which form a super-set to the E+D architecture at the cost of structural complexity.

An important finding in our analyses concerns the asymmetric roles played by the encoder and the decoder, positioned on the same plane. Although it may be tempting to view the encoder as an input port and the decoder as the sole site of learned functional computation, our bias-augmented construction shows that the encoder region itself carries unexploited degrees of freedom that, when activated through a trained/frozen phase bias $b(x)$, systematically improve the nonlinear function approximation fidelity on both randomly sampled target nonlinear function ensembles (**Fig. 4**) and structured nonlinear activation functions of practical interest (**Fig. 5**). Because the phase bias terms, $b(x)$, is optimized jointly with the decoder profile under the same end-to-end training loss, the encoder and decoder co-adapt rather than specialize sequentially, and the net effect is an enlargement of the accessible solution space without any changes in the architectural depth, the readout complexity, or the number of diffractive surfaces. The target-specific phase profiles visualized in **Fig. 5b** make this cooperation tangible: at a fixed input $a$, the encoder regions differ across target functions precisely because the optimized/frozen $b(x)$ has absorbed some function-specific structure that was otherwise forced to live entirely in the decoder space.

As demonstrated in our experimental results, practical realization of the E+D processor is naturally suited to one SLM operating on the input plane, with the decoder region hosted either as a static/fixed modulation region of the same SLM or as a fabricated/frozen phase plate in the same plane. Because SLMs impose a finite phase bit depth, our demonstrations that the trained encoder bias sustains — and in fact widens — its performance advantage under coarse input-plane phase quantization (**Fig. 8**) is a prerequisite for experimental implementations under limited phase bit-depth. The mechanism underlying this robustness is that the trained per-pixel offsets in the frozen bias term $b(x)$ desynchronize the quantization thresholds seen by different encoder pixels, so that the aggregate encoder field responds more continuously to small variations in $a$ than the strictly phase-locked baseline allows. This is closely analogous to dithering in digital signal conversion, where deliberate perturbations decorrelate quantization artifacts; here, the perturbations are learned rather than stochastic, and they are folded into the E+D architecture at no additional hardware cost. We expect this mechanism to generalize beyond the specific bit-depth range studied here, and to be particularly valuable for SLM platforms such as phase-only devices with 6–8-bit phase control, where naive implementations of the bias-free diffractive encoder would otherwise suffer performance artefacts.

Several extensions of the current work are worth noting. First, the theoretical analysis and the universal-approximation claim apply to real-valued scalar input variables and to band-limited real-valued target functions; extending the framework to multi-variable inputs or to vector-valued targets (e.g., by partitioning the output plane into multiple detectors, each synthesizing a distinct functional component) is a natural next step that the presented E+D architecture accommodates without structural changes. Second, although our simulations and experimental demonstration used monochromatic illumination, the E+D architecture extends naturally to wavelength-multiplexed operation, where each spectral channel can independently realize an E+D processor on the same surface. This direction — together with massively parallel multi-pixel output intensity detection — would enable the simultaneous synthesis of a large number of nonlinear functions using a single E+D surface.

Finally, we anticipate that the architectural simplicity of E+D processors will enable compact and robust optical implementations of nonlinear analog computation that are practical to fabricate, align, and deploy at scale, and that the principles established here will inform the design of future diffractive computing systems — including those operating in higher input dimensions, under incoherent or partially coherent illumination, and across multiple spectral channels.

## Material and methods

### Experimental set-up

The coherent E+D processor is experimentally implemented using a visible-light phase-modulation set-up. A 635 nm coherent laser source (Fianium) was used to illuminate a reflective phase-only SLM (Holoeye PLUTO-2.1; resolution, $1920 \times 1080$; pixel pitch, 8 μm). The SLM is operated with 8-bit phase modulation. The modulated optical field is relayed through a 4f optical system and captured by a CMOS camera (Basler GigE acA1920-40gm; resolution, $1920 \times 1200$; pixel pitch, $5.86$ $\mu$m). The 4f system is used only as a relay optics, without applying Fourier-plane filtering or modulation.

For the parallel, spatially-multiplexed nonlinear-function approximation experiment, nine function-specific E+D units are displayed simultaneously on the SLM in a $3 \times 3$ spatial mosaic. Each E+D unit occupies a $1.6\ mm \times 1.6\ mm$ ($200 \times 200$ SLM pixels) region on the SLM. The scalar input argument $a$ is encoded into a phase profile over an encoder region, while the remaining pixels in each tile formed the trainable decoder region. The target functions are generated using a random Fourier-series representation with $N_p = 4$ harmonic channels. Accordingly, the native encoder consisted of a $2 \times 2$ set of phase-encoded pixels. To increase the optical power collected from the encoder region and improve the signal-to-noise ratio of the measured response, each native encoder pixel was expanded into a $0.2\ mm \times 0.2\ mm$ ($25 \times 25$ SLM pixels) macro-pixel on the SLM, yielding a $0.4\ mm \times 0.4\ mm$ ($50 \times 50$ SLM pixels) encoder region for each E+D processor unit.

The nine E+D units simultaneously produce nine output responses on the camera plane. For each target nonlinear function, a corresponding $0.4\ mm \times 0.4\ mm$ ($68 \times 68$ detector pixels) detector region is defined on the CMOS image sensor. The measured optical output for each function is obtained by integrating the camera intensity within its assigned detector region. The resulting detector outputs are used to evaluate the experimentally realized nonlinear functions – performed all in parallel.

### Supplementary Information includes:

- Physical forward model and training details
- Training and testing set-up

- Supplementary Figures S1-S5

## Figures

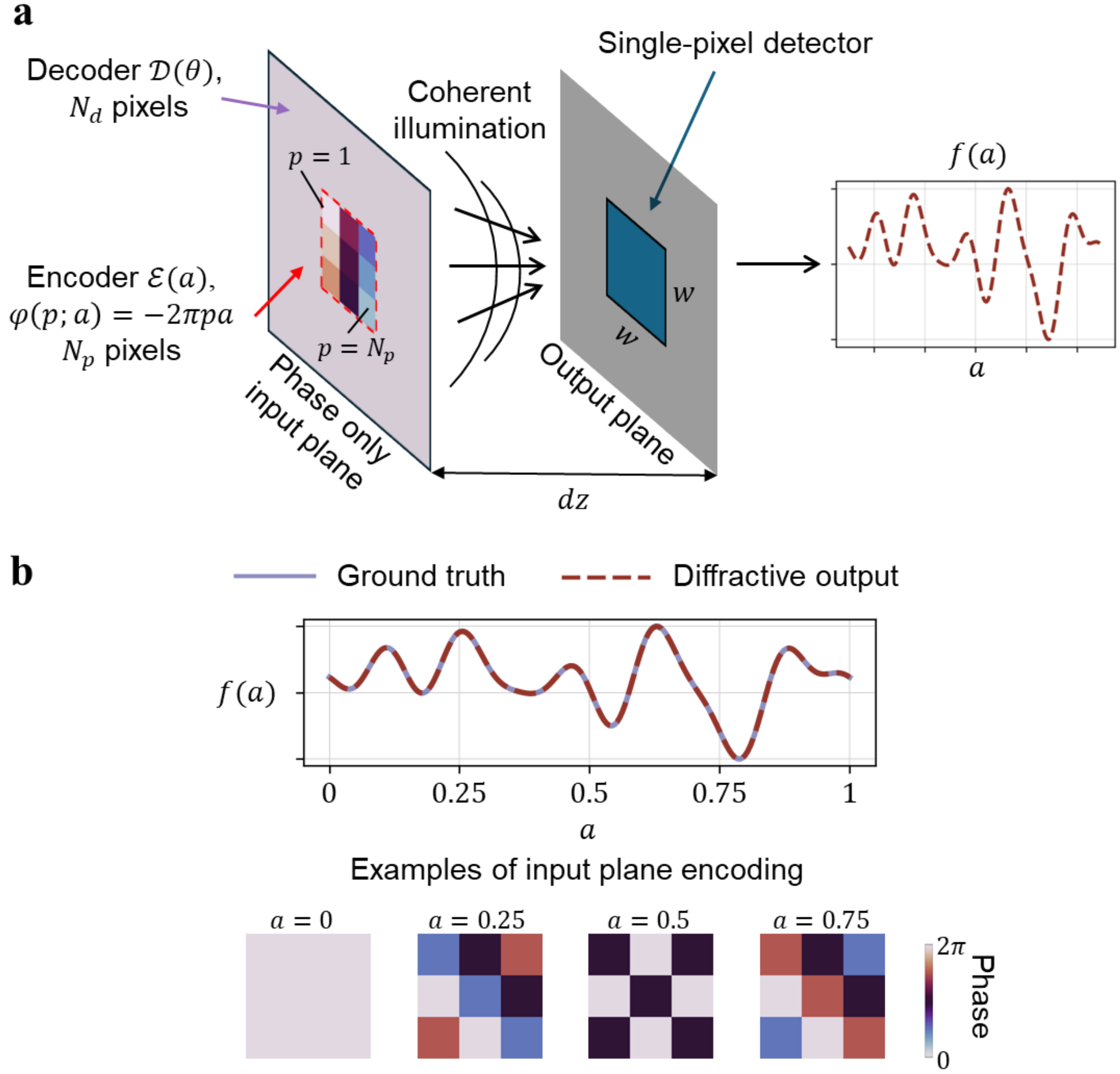

**Fig. 1. Single-layer diffractive optical processor (E+D processor) for nonlinear function approximation under coherent illumination. a** Schematic of the proposed architecture. A phase-only input plane with uniform intensity illumination is partitioned into $\mathcal{E}(a)$, an argument-dependent encoder containing $N_p$ pixels and $\mathcal{D}(\theta)$, a static optimizable decoder containing $N_d$ pixels (frozen after optimization). The encoder region projects the scalar input $a$ into the phase profile, being co-localized with the optimized decoder region on the same diffractive surface; the resulting optical fields from both regions propagate an identical distance $dz$ to the output plane, where a single-pixel optical detector of width $w$ integrates the optical intensity to generate an

estimate of the target nonlinear function, $f(a)$. **b** Representative example of nonlinear function approximation by the E+D optical processor, showing the well alignment of the ground-truth function (solid curve) and the diffractive output (dashed curve). Examples of encoder phase patterns for selected values of $a$ are also illustrated at the lower panel.

**a**

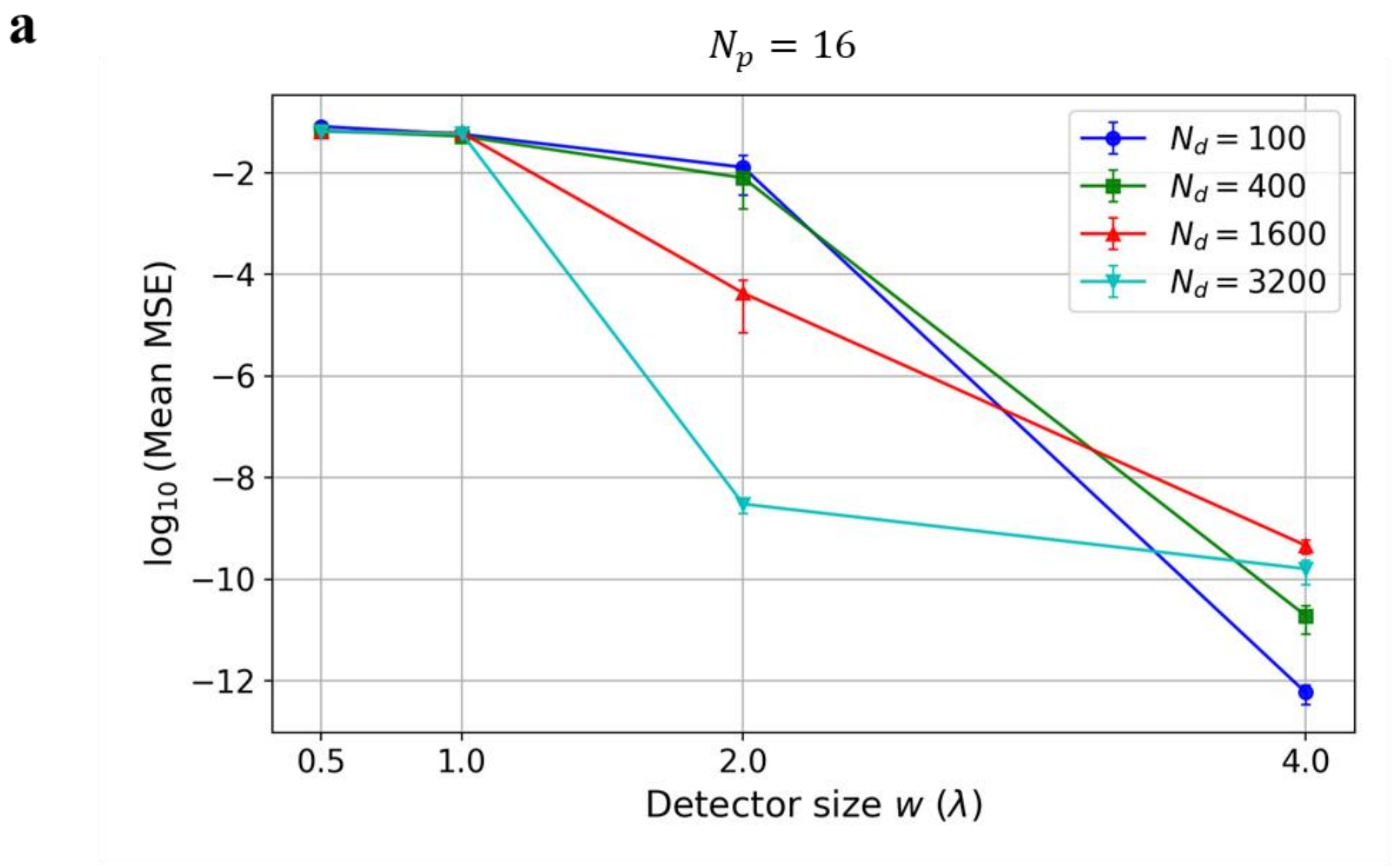

**b**

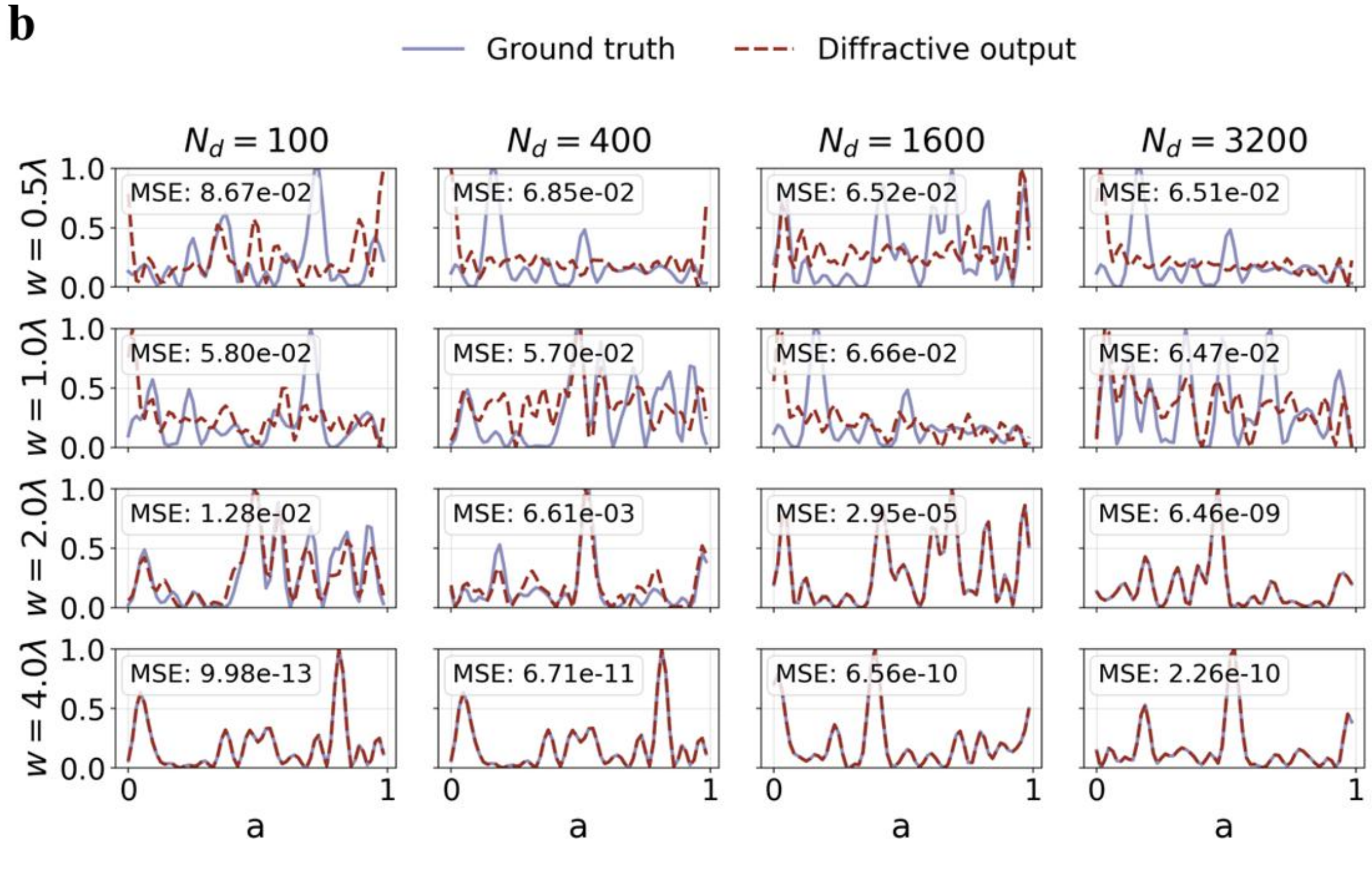

**Fig. 2. Nonlinear function approximation performance as a function of the detector size ($w$) and the number of trainable pixels in decoder-region ($N_d$). a** Mean squared error (MSE) of the diffractive output versus the detector width $w$ for different decoder region sizes $N_d$; target

nonlinear functions are composed of $N_p = 16$ Fourier harmonics. Data points represent the mean MSE across 20 independently trained E+D processor models, with error bars indicating the standard deviation across different runs. **b** Representative function-approximation results for different combinations of $w$ and $N_d$. Solid curves denote the ground truth nonlinear functions and dashed curves denote the corresponding diffractive outputs.

**a**

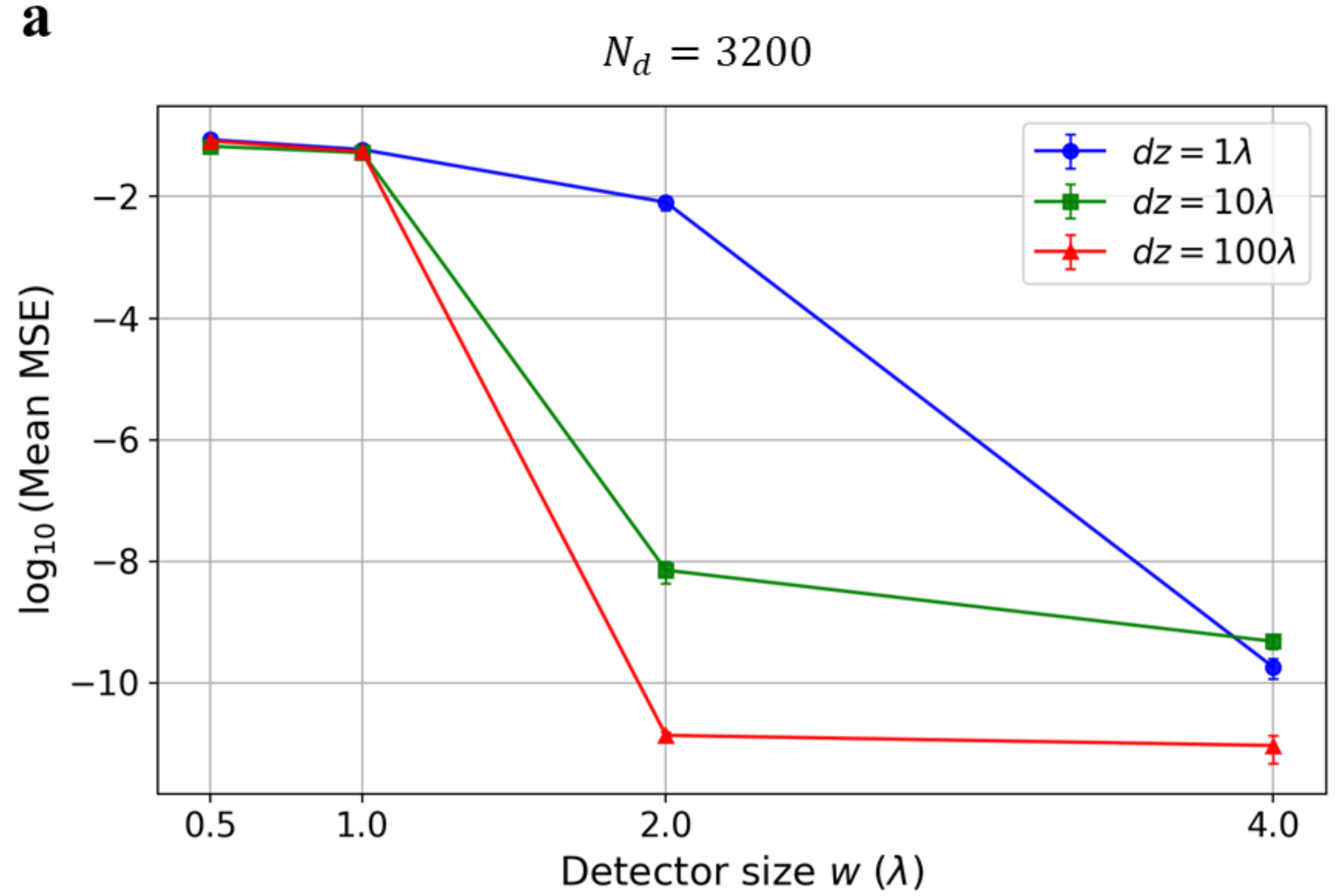

**b**

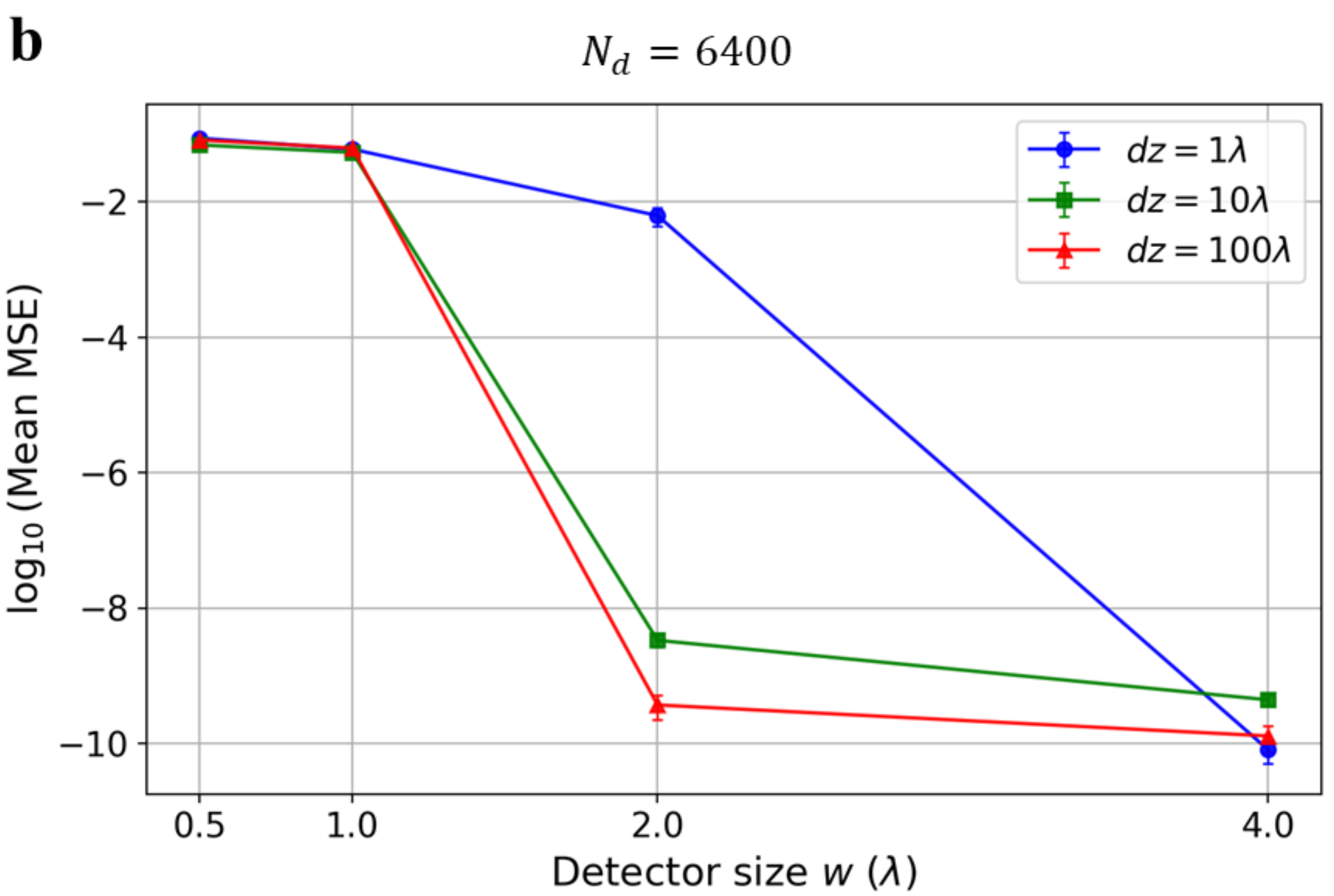

**Fig. 3 Nonlinear function approximation performance of E+D processors as a function of detector size ($w$) and propagation distance ($dz$).** The MSE of the diffractive output versus the detector width $w$ for different axial propagation distances $dz$; target nonlinear functions are

composed of $N_p = 16$ Fourier harmonics. Data points represent the mean MSE across 20 independently trained E+D processor models, with error bars indicating the standard deviation across differen runs. Results are presented for decoder region sizes of **a** $N_d = 3200$ and **b** $N_d = 6400$.

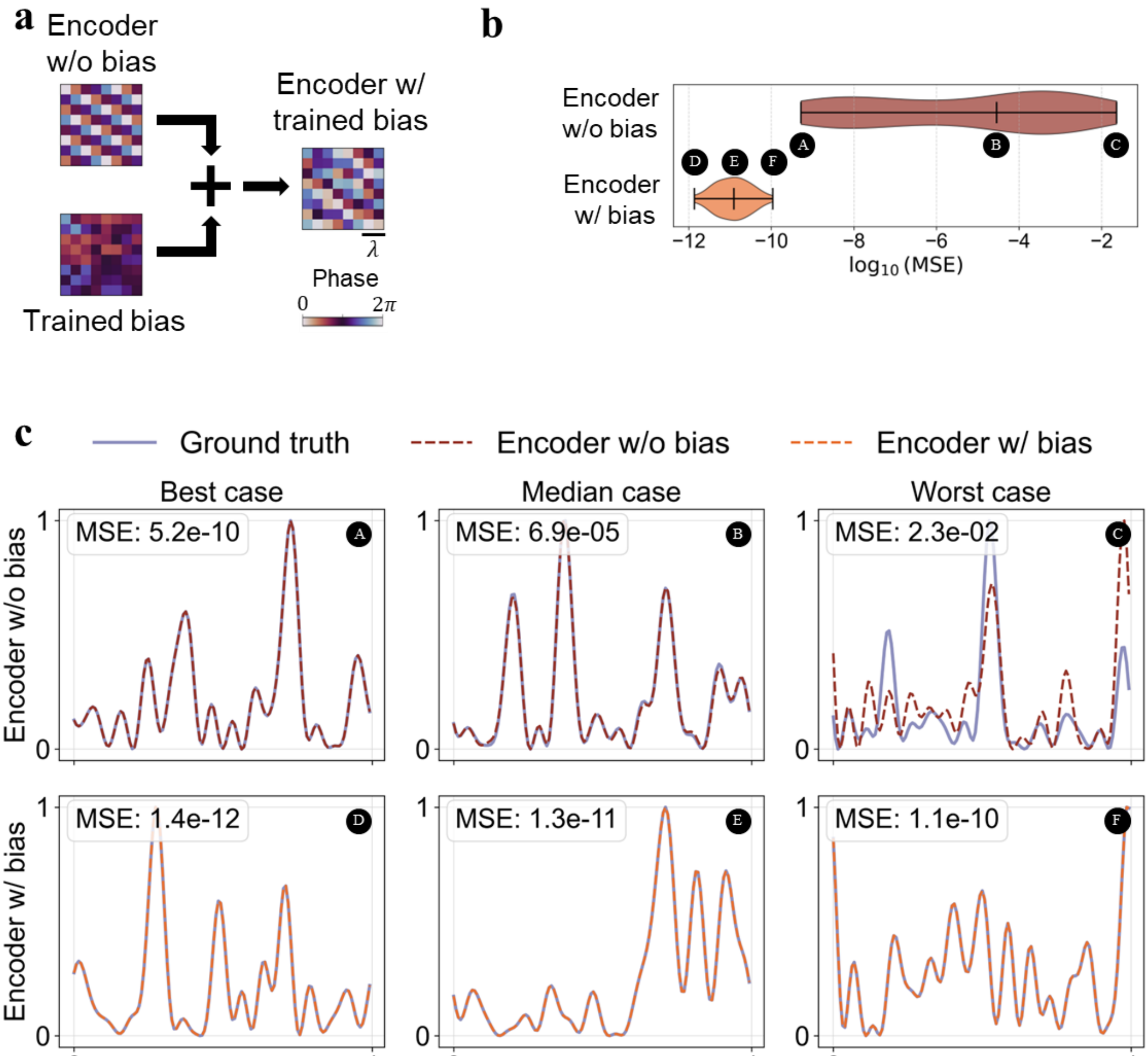

**Fig. 4. Nonlinear function approximation improvement enabled by incooperating an encoder region frozen phase bias. a** Schematic of the bias-augmented encoder. A trained/frozen phase bias is added to the argument-dependent encoder pattern, yielding a modified encoder phase profile that is used during inference. **b** Approximation error distribution comparison of the same set of nonlinear functions with and without the frozen phase bias. Each distribution is compiled from 40 independently sampled target functions, each optimized with a separately trained E+D processor model. **c** Representative approximation results without (top row) and with (bottom row) the trained encoder bias, showing the best-, median- and worst-case examples. Solid curves denote the ground truth nonlinear functions, dashed brown curves denote the outputs without phase bias, and dashed orange curves denote the outputs with a trained/frozen phase bias.

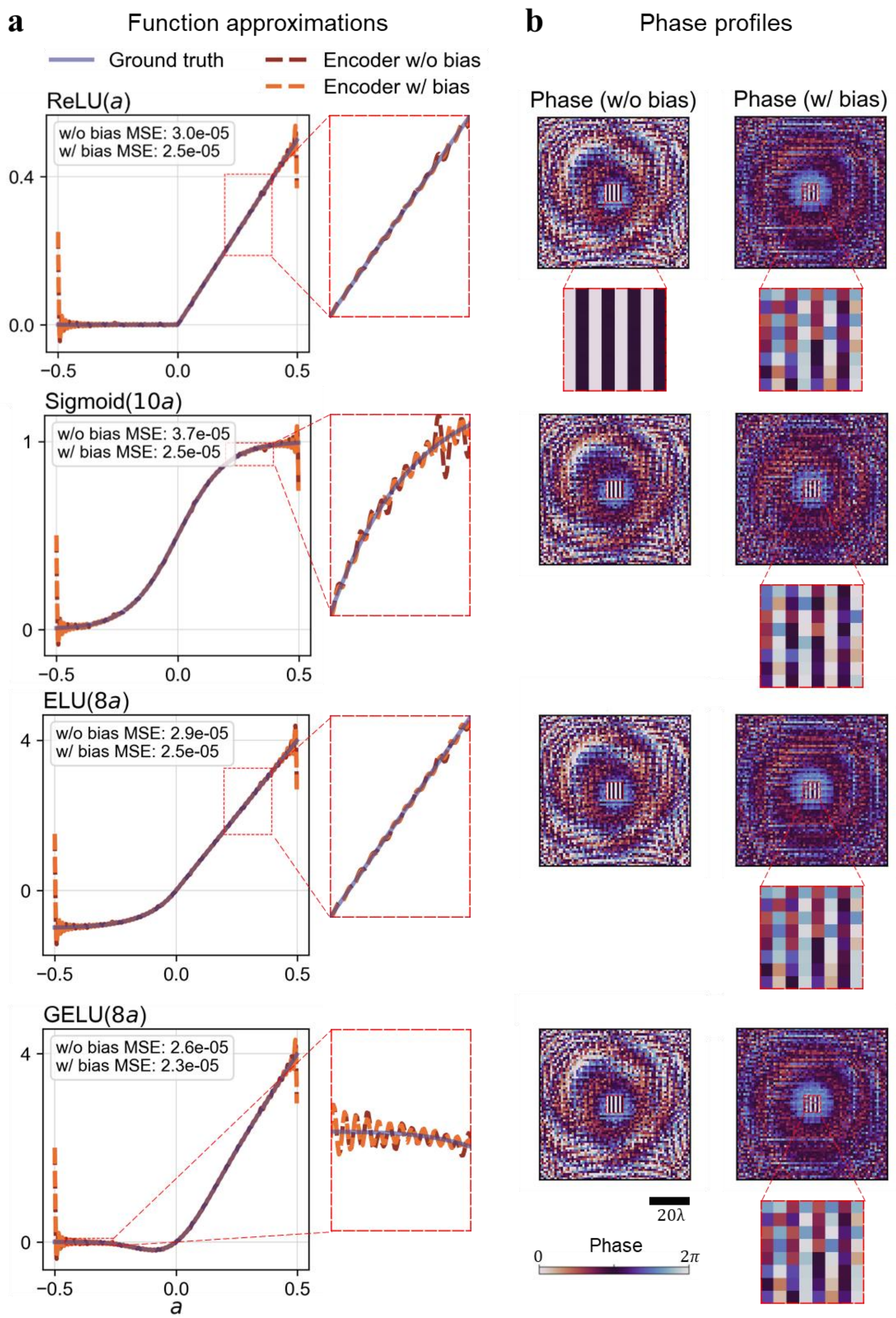

**Fig. 5. Comparison of the approximation of representative nonlinear activation functions with or without phase bias. a** Approximation results for four representative nonlinear activation

functions commonly used in artifical neural networks, comparing the ground truth (solid curves), the diffractive outputs obtained without an encoder bias (brown dashed curves), and diffractive outputs obtained with a trained/frozed encoder bias (orange dashed curves). To provide a fair comparison, the reported MSE values are computed after normalizing each target function to the range $[0, 1]$; the curves are then rescaled to their original ranges for visualization. To reduce boundary artifacts caused by Gibbs phenomenon, the MSE is evaluated over $a \in [-0.5 + \epsilon,\ 0.5 - \epsilon]$ with $\epsilon = 2.5\%$. **b** Optimized phase profiles of the E+D optical processor without and with the trained encoder bias, at a representative input $a = 0.5$, for each of the target nonlinear activation functions.

**a**

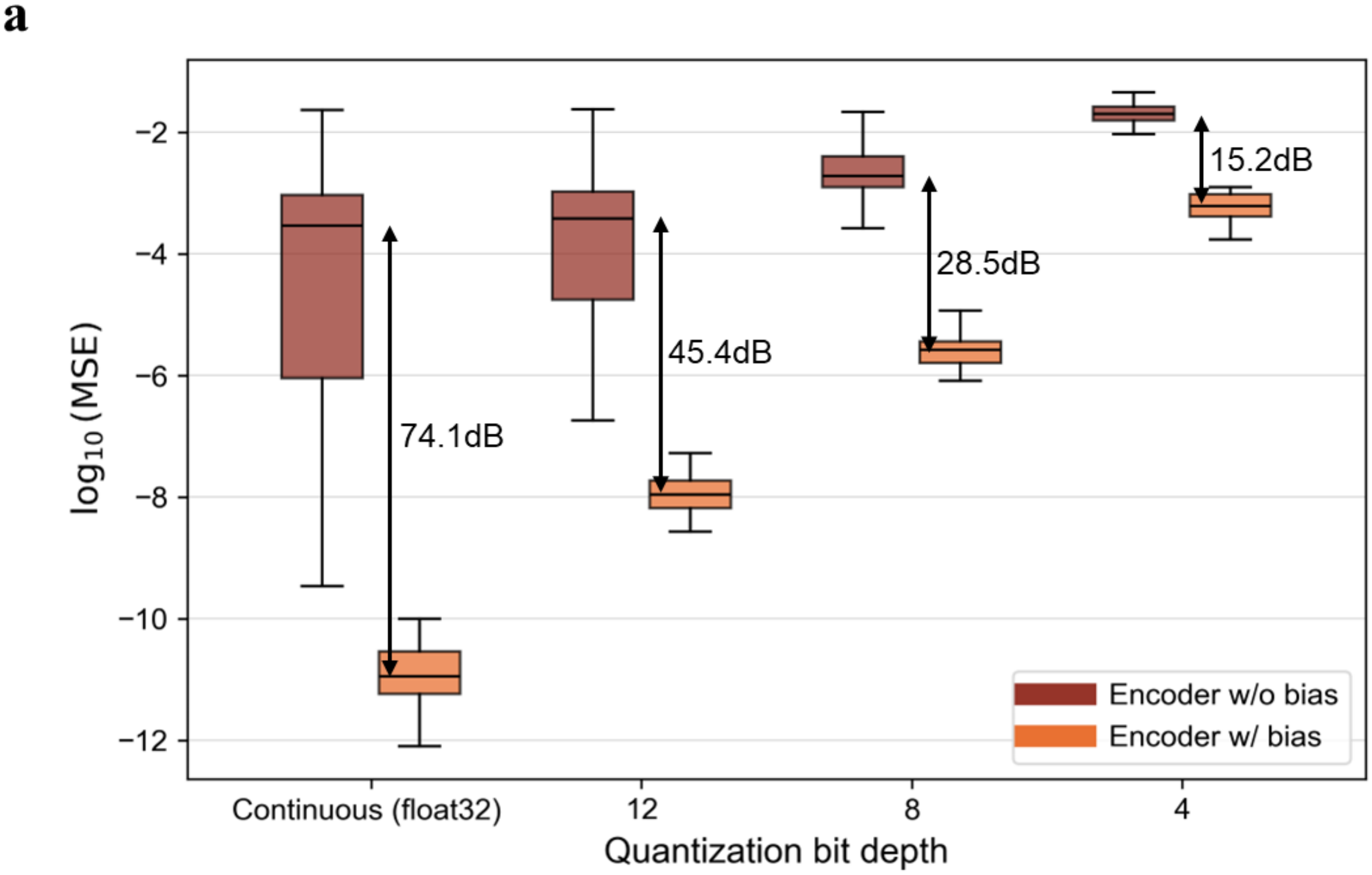

**b**

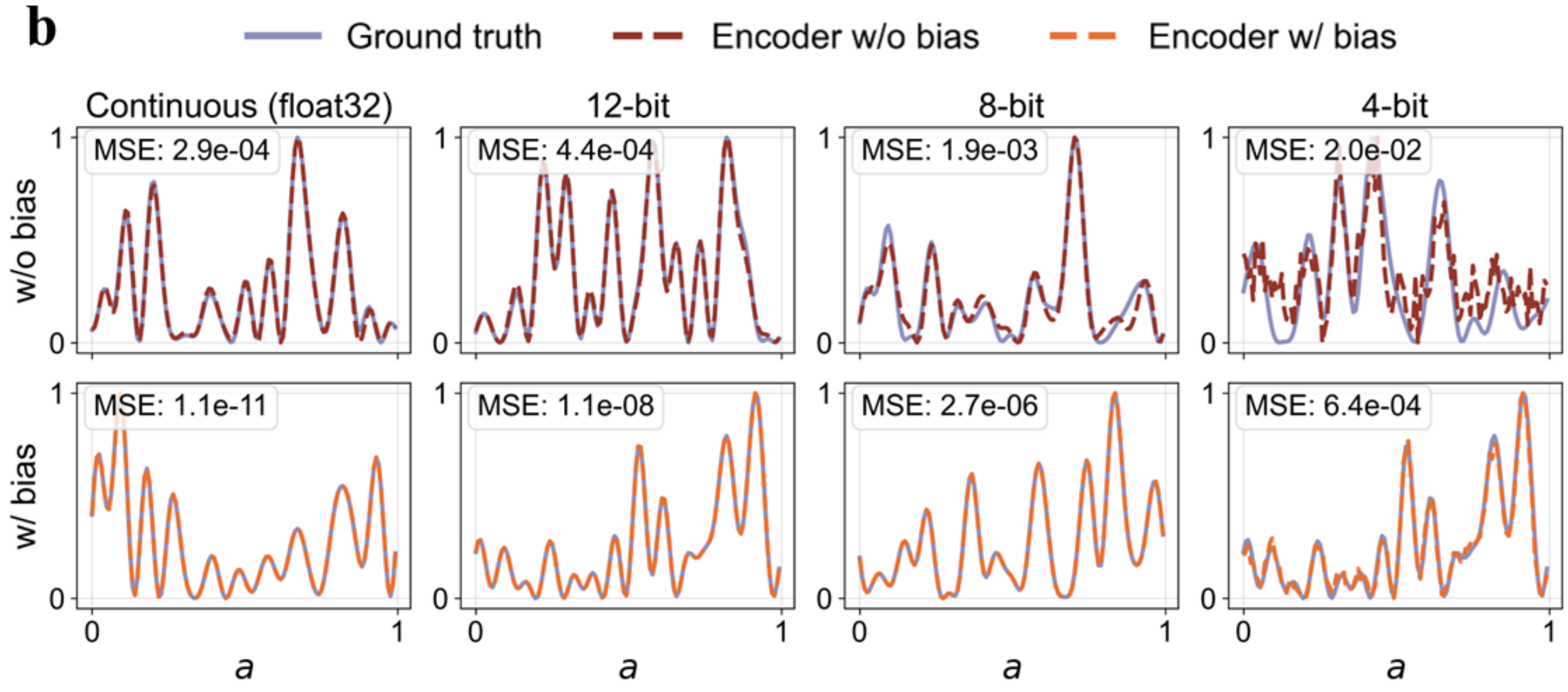

**Fig. 6. Function approximation performance of the E+D processor with and without a trained encoder bias under limited phase bit depth. a** Approximation error distribution comparison of the same set of nonlinear functions with and without a trained phase bias under different quantization bit depth. The trained/frozen bias enhances robustness by lowering errors across all quantization levels. The continuous benchmark uses unquantized float32 phase values (~24-bit precision), which is used only for comparison purposes. **b** Representative approximation

results under different encoder quantization levels. Solid curves denote the ground truth nonlinear functions, brown dashed curves denote the outputs obtained without the trained bias, and orange dashed curves denote the outputs obtained with the trained/frozen bias.

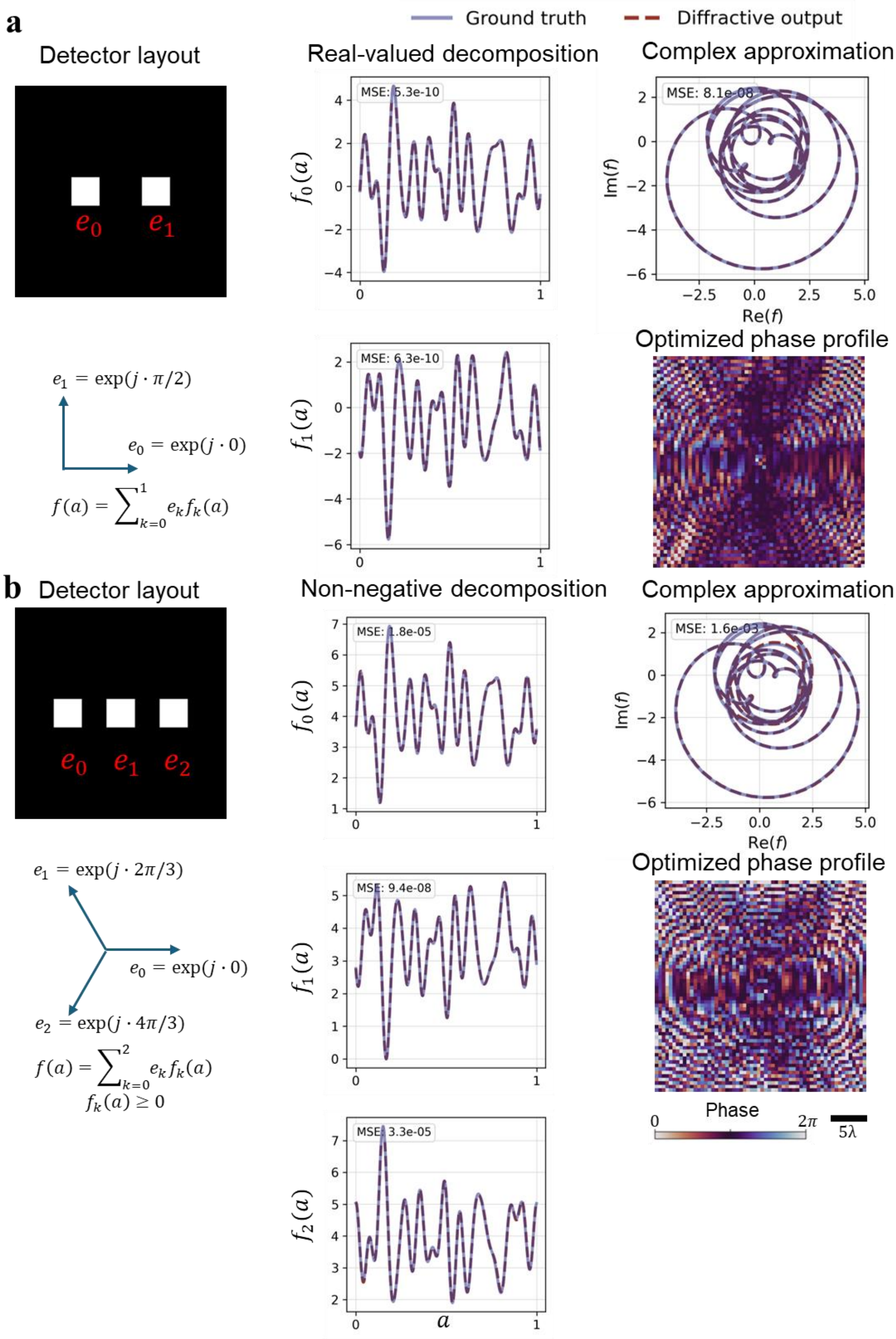

**Fig. 7 Complex-valued nonlinear function approximation using E+D processors. a**, Two-detector scheme for approximating a complex-valued nonlinear function using an E+D processor.

Two separated output detectors are assigned to two distinct real-valued nonlinear functions, $f_0(a)$ and $f_1(a)$, which are combined with two orthogonal complex bases $e_0$ and $e_1$, to form $f(a) = \sum_{k=0}^{1} e_k\, f_k(a)$. The left panels show the detector layout and basis vectors, the middle panels compare the ground-truth nonlinear functions, $f_0(a)$ and $f_1(a)$, with the diffractive outputs, and the right panels show the reconstructed complex-valued function trajectory and the optimized E+D phase profile. **b** Three-detector scheme using non-negative real-valued nonlinear functions. Three separated detectors approximate distinct non-negative real-valued functions $f_k(a) \geq 0$, $k = 0,1,2$, which are combined with bases $e_k = \exp(j2\pi k/3)$ to reconstruct the target complex-valued function, avoiding the need for signed detector outputs or background removal.

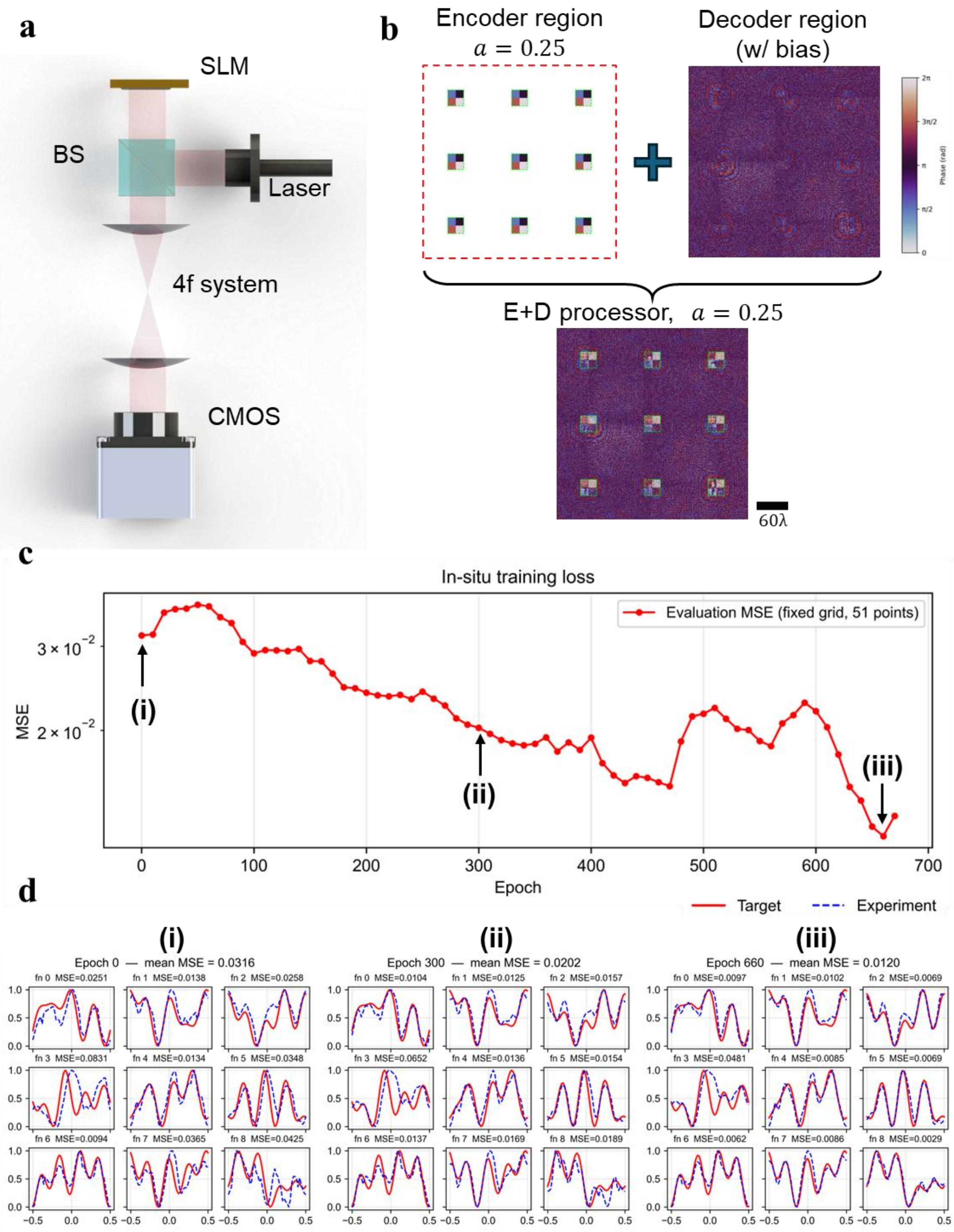

**Fig. 8. Experimental demonstration of a coherent E+D processor, simultanously approximating $N_f = 9$ functions. a** Schematic of the experimental set-up. A coherent laser beam

is redirected by a beam splitter (BS) before modulated by an SLM. The reflected light propagates through a 4f system, which acts as a simple relay system, and is captured by a CMOS camera. **b** Representative phase patterns displayed on the SLM for an input argument $a = 0.25$ at epoch 660. The argument-dependent encoder region is combined with a trained decoder region, including a learned phase bias, to form the full E+D processor used during inference. **c** Evaluation MSE evolution during in situ training of the experimental E+D optical processor, calculated using 51 uniformly disrtributed input argument $a \in [0,1]$ . **d** Representative nonlinear-function approximation results at epochs 0, 300 and 660, showing the progressive improvement enabled by model-free in situ training. Solid curves indicate the ground-truth nonlinear functions, while dashed curves represent the experimentally measured outputs.